\documentclass[preprintnumbers,amsmath,amssymb,aps,twocolumn]{revtex4}
\usepackage{graphicx}

\newcommand{\smallstep}{\vspace{.08em}}

\def\di{\displaystyle}

\def\bg{\begin{eqnarray}\begin{array}{rcl}\displaystyle}
\def\eg{\end{array} &\di    &\di   \end{eqnarray}}
\def\bm#1{\begin{eqnarray}\begin{array}{#1}\di}
\def\bmo#1{\begin{eqnarray*}\begin{array}{#1}\di}
\def\bml#1#2{\begin{eqnarray}\begin{array}{#1}\label{#2}\di}
\def\bgo{\begin{eqnarray*}\begin{array}{rcl}\displaystyle}
\def\ego{\end{array} &\di    &\di \nonumber  \end{eqnarray*}}

\def\btensor#1#2{\renew\left#1\begin{array}{#2}\di}
\def\brtensor#1#2#3{\ren#3\left#1\begin{array}{#2}}
\def\botensor#1#2{\renew\left#1\begin{array}{#2}}
\def\etensor#1{\end{array}\right#1}

\def\eq#1{(\ref{#1})}
\def\Eq#1{Eq.~(\ref{#1})}

\def\tr{{\rm tr}}
\def\Tr{{\rm Tr}}

\def\s0#1#2{\mbox{\small{$ \frac{#1}{#2} $}}}
\def\0#1#2{\frac{#1}{#2}}

\def\CF{{\mathcal F}}

\def\ren#1{\renewcommand{\arraystretch}{#1}}

\def\renew{\renewcommand{\arraystretch}{1}}

\begin{document}

\title{On the Yang-Mills two-loop effective action with wordline methods}
\vspace{1.5 true cm}

\author{Jan M.~Pawlowski${}^{a}$} 
\author{Michael G.~Schmidt${}^{a}$}
\author{Jian-Hui~Zhang${}^{b}$} 
\address{ ${}^a$\mbox{Institut
      f{\"u}r Theoretische Physik, Universit\"at Heidelberg,
      Philosophenweg 16, 69120 Heidelberg,
      Germany}\\
    ${}^b$\mbox{Max-Planck-Institut f{\"u}r Physik,
     F\"ohringer Ring 6, 80805 M\"unchen,
      Germany}}


\thispagestyle{empty}

\abstract{ We derive the two-loop effective action for covariantly
  constant field strength of pure Yang-Mills theory in the presence of
  an infrared scale. The computation is done in the framework of the
  worldline formalism, based on a generalization procedure of
  constructing multiloop effective actions in terms of the bosonic
  worldline path integral. The two-loop $\beta$-function is correctly
  reproduced. This is the first derivation in the worldline
  formulation, and serves as a nontrivial check on the consistency of
  the multiloop generalization procedure in the worldline formalism.}}

\maketitle

\pagestyle{plain}
\setcounter{page}{1}

\section
{Introduction}

The physics of strong but slowly varying chromomagnetic and electric
fields may provide some insight to the non-trivial vacuum structure of
QCD.  For example, the Euler-Heisenberg Lagrangian in QED exhibits
truly non-perturbative effects; it allows the investigation of the
non-linear regime of QED, and has also been studied beyond one loop,
see e.g.\ \cite{Ritus:1977iu,Dittrich:1985yb,Dunne:2001pp}. In QCD
non-linearities already are present on the classical level due to its
non-Abelian nature. It is yet unknown how the full effective action
changes beyond one loop. The computation of multiloop terms in the
effective action is cumbersome, in particular for a non-Abelian gauge
group. These computations simplify if background field methods are
employed and the multiloop terms are evaluated for specifically chosen
background configurations such as covariantly constant or selfdual
configurations. This gives access to beta functions and parts of the
effective action beyond one loop, see e.g.\
\cite{Caswell:1974gg,Abbott:1980hw,Bornsen:2002hh}. Wordline methods
\cite{Bern:1991aq,Strassler:1992zr,Schmidt:1993rk,Schmidt:1994zj,%
  Schmidt:1994aq,Reuter:1996zm,Sato:1998sf,Sato:2000cr,Gies:2001zp,%
  Schmidt:2003bf,Schubert:2001he,Kors:2000bb,Magnea:2004ai,Dai:2008bh}
have been shown to lead to a striking simplification of certain
computations. A summation of Feynman diagrams is already implemented
without loop momentum integrals and Dirac traces
\cite{Schmidt:1994zj,Schmidt:1994aq}.  \smallstep

In the present work we provide for the first time a numerically
accessible expression for the two loop effective action of Yang-Mills
theory for covariantly constant fields in the presence of a physical
infrared cut-off (at finite correlation length). This computation
serves many purposes: firstly, it completes the construction of
multiloop worldline methods for Yang-Mills theories as initiated in
\cite{Sato:1998sf,Sato:2000cr}. The two-loop beta function serves as a
non-trivial consistency check. Secondly, it is the necessary input for
an RG-improved non-perturbative computation of the effective action
within a Wilsonian framework. Finally one can study the stabilisation
of the Savvidy vacuum beyond one loop. \smallstep

\section {wordline representation beyond one loop}

We briefly recapitulate the analysis of
\cite{Sato:1998sf,Sato:2000cr}. The starting point is the generating
functional of pure Yang-Mills theory in the presence of a background
field configuration $A$,
\begin{eqnarray}\label{eq:Z}
  Z[A]=\int D a\, \exp -S[a,A]\,, 
\end{eqnarray} 
where the gauge-fixed Yang-Mills action is given by
\begin{eqnarray}\label{eq:S}
  S[a,A]&=&\012 \int_x \tr\, F_{\mu\nu}(a+A)
  ^2 \\\nonumber 
  && -\0{1}{2 \xi}\int_x \tr\, (Da)^2 -\Tr\,\ln (-D(A) D(a+A))\,,
\end{eqnarray} 
with $\tr\, t^a t^b =- \delta^{ab}/2$ in the fundamental
representation, $D(A)=\partial-i g A$, and $g F_{\mu\nu}=i [D_\mu,
D_\nu]$. Within a perturbative expansion $Z[A]$ reads
\begin{eqnarray}\label{eq:pertexp}
  Z[A]=\sum_n Z_n[A]\,,
\end{eqnarray} 
where $Z_n$ comprises the $n$-loop contribution to the generating
functional. The generating functional \eq{eq:pertexp} is gauge
invariant under the gauge transformation $A\to A+D\omega$, and its
logarithm is the gauge invariant Wilsonian effective action of pure
Yang-Mills. The one loop contribution $Z_1[A]$ for a subset of field
configurations, e.g.\ covariantly constant fieldstrength $F$, has been
computed in \cite{Reuter:1996zm}, for related results within standard
methods see \cite{Shore:1981mj}.  So far, a full computation of the two loop
contribution $Z_2[A]$ is lacking.  After some algebra, \eq{eq:Z} can
be turned into a more convenient representation for $Z_2[A]$, see 
\cite{Sato:1998sf}.  However, in all representations the formal
expression \eq{eq:Z} suffers both from UV and IR divergences.  In the
present work we regularise and renormalise these divergences
separately: for the UV divergences we employ dimensional
regularisation for analytic purposes and construct a gauge-invariant
proper-time cutoff interesting for numerical work. The divergences are
then cured by appropriate counter terms.  Additionally we introduce a
physical IR cut-off. IR divergences are absent, if putting the theory
into a box of size $L$.  Effectively this can be implemented by
introducing gauge invariant masses $m\sim 1/L$ to the propagating
degrees of freedom of the theory. The latter also has the advantage of
implementing a physical mass gap on the level of the Green functions.
This offers a path towards a self-consistent investigation of QCD in
the confining regime in an effective field theory approach that is
quite close to the fundamental theory. We emphasise that the approach
leads to a fully gauge invariant effective action. However, the choice
of the gauge-fixing parameter is directly related to the choice of
different {\it physical} boundary conditions on the surface of the
box.   \smallstep

Within this framework the renormalised two loop contribution $Z_2[A]$
is provided by \cite{Sato:1998sf}
\begin{widetext}
\begin{eqnarray}\nonumber 
Z_2[A] &=& \exp\Bigl\{ \frac{i}{2}
  \int_{y_1,y_2}\Bigl(-\frac{i}{2}
  \frac{\delta}{\delta\alpha_{\mu\nu}^a(y_1)}D_\mu^{ai}(y_1)
  +i\frac{\delta}{\delta\beta_\nu^i(y_1)}\Bigr) 
  \bar\Delta_{\nu\sigma}^{ij}(y_1,y_2)\Bigl(-\frac{i}{2}
  \frac{\delta}{\delta
    \alpha_{\rho\sigma}^e(y_2)}D_\rho^{ej}(y_2)
+i
  \frac{\delta}{\delta\beta_\sigma^j(y_2)}\Bigr)\Bigr\}\\[2ex]
&&\hspace{.3cm} \times  \exp\{\frac{i}{16}
    \frac{\delta}{\delta\alpha_{\mu\nu}^a}1
    \frac{\delta}{\delta\alpha^{a,\mu\nu}}\} \exp\{
    -\frac{1}{2}\Tr\ln\bar\Delta^{-1}\} \left. \exp\{\Tr\ln 
      (D^2+m^2-D\beta f)\}\right|_{\alpha=\beta=0}+
{\rm c.t.}\,,
\label{eq:derZ}\end{eqnarray}
\end{widetext}
with antisymmetric tensor field $\alpha_{\mu\nu}=-\alpha_{\nu\mu}$, 
\begin{eqnarray}
  (\bar\Delta^{-1})_{\mu\nu}^{ab}=(\Delta^{-1}_{\mu\nu})^{ab}
  +2f^{abc}\alpha_{\mu\nu}^c\,,
\end{eqnarray}
and
\begin{eqnarray}\label{eq:prop} 
  \Delta_{\mu\nu}^{ab}=[-g_{\mu\nu}D^2+
  2F_{\mu\nu}^cf^{abc}+m^2+(1-\frac{1}{\xi})D_\mu D_\nu]^{-1}\,.
\end{eqnarray} 
We emphasise again that $m^2$ serves a twofold though related purpose.
Firstly it accounts for a possible non-perturbative mass-gap, secondly
it mimics the implementation of a finite volume.  The space-time
integration over $y_1,y_2$ can be used for regularising the
generating functional \eq{eq:derZ} by means of the dimensional
regularisation
\begin{eqnarray*}
\int_y =\int d^D y\,,\qquad {\rm with} \qquad D=4 -2 \epsilon\,. 
\end{eqnarray*} 
In Feynman gauge $\xi=1$, the expression simplifies as the last term in
\eq{eq:prop} drops out, and the non-trivial tensor structure
disappears.  In \eq{eq:derZ} we have not specified the counter terms indicated by ${\rm c.t.}$, that shall be discussed later.  Within
the representation introduced above the effective action derived from
the generating functional \eq{eq:derZ} reads
\begin{eqnarray}
  \Gamma=\Gamma_{\rm gluon}+\Gamma_{\rm ghost}\,,
\end{eqnarray}
with
\begin{eqnarray}\nonumber 
  &&\hspace{-.5cm}  \Gamma_{\rm ghost}= -\012 \int_{y_1,y_2}
  \0{\delta}{\delta\beta}\bar\Delta\0{\delta}{\delta\beta} \Tr\ln 
  (D^2+m^2-D\beta f) 
  |_{\beta=0}\\ 
  && \hspace{1.5cm}+{\rm c.t.}\,,\label{eq:ghostGa}\end{eqnarray}
the contribution of purely gluonic loops, $\Gamma_{\rm gluon}$, has been 
computed in \cite{Sato:2000cr}, and the result is summarised in
Appendix~\ref{app:gluon}. We complete the analysis of \cite{Sato:2000cr} by
computing the ghost contribution $\Gamma_{\rm ghost}$ as well as the
renormalisation insertions.  To that end we turn \eq{eq:ghostGa} into Euclidean
wordline integrals, and arrive at \cite{Sato:1998sf}
\begin{widetext}
\begin{eqnarray}\nonumber 
\Gamma_{\rm ghost} &=& -\frac{1}{2}\int_{y_1,y_2}\int_{\tau}^{\infty}
  dT_1dT_2dT_3\int^{x(T_1)=y_2'}_{x(0)=y_1}[\mathcal{D}x]_{T_1}
\int^{\bar x(T_2)=y_1'}_{\bar x(0)=y_2}[\mathcal{D}\bar x]_{T_2}
  \int^{w(T_3)=x(T_1)}_{ w(0)=\bar x(T_2)}[\mathcal{D}w]_{T_3} \\[2ex] 
& &
\hspace{2cm} \times  (\mbox{Pe}^{\int_0^{T_3}M(w)})_{\mu\nu}^{ae}
  (\mbox{Pe}^{\int_0^{T_1}N(x)})^{cj}\stackrel{\leftarrow}{D}
  \!_\nu^{fj}(y_2')(\lambda^e)_{fg}
 (\mbox{Pe}^{
    \int_0^{T_2}N(\bar x)})^{gi}\stackrel{\leftarrow}{D}\!_\mu^{bi}(y_1')
  (\lambda^a)_{bc}|_{y_i'=y_i} +{\rm c.t.}\,,
\label{eq:preGaghost}\end{eqnarray}
\end{widetext} 
with the abbreviations
\begin{eqnarray}\nonumber 
  \int[\mathcal{D}x]_T F[x]&=&\int\mathcal{D}x\,
  e^{-\int_0^T d\tau( \frac{1}{4}\dot x^2(\tau)+m^2)}F[x]\nonumber\,,
\end{eqnarray}
and 
\begin{eqnarray}\nonumber
  M_{\mu\nu}^{ab}[x(\tau)]&=&2i(F_{\mu\nu}^c-\delta_{\mu\nu}
  \frac{1}{2}A_\eta^c\dot x^\eta)(\lambda^c)^{ab}\,,\\
  N(x)^{ab}&=&-iA_\mu^c\dot x^\mu(\lambda^c)^{ab}\,.
\end{eqnarray} 
\Eq{eq:preGaghost} stands for the fully renormalised ghost
contribution to the two loop effective action. We have employed an
additional ultraviolet regularisation in the proper-time integrations
which entails a gauge invariant momentum cut-off. Such a cut-off
scheme is amiable to numerical computation, whereas the dimensional
regularisation facilitates analytic computations. In the following we
shall conveniently project onto either regularisation by simply
switching off either the regularisation parameter $\tau\to 0$ or $\epsilon\to 0$. We have not
specified the counter terms indicated by ${\rm c.t.}$, which in general
depend both on the proper-time regularisation via $\tau$ and on the
dimensional regularisation via $\epsilon$. The
computations of these counter terms will be discussed in the next
section.

For explicit computations we employ pseudo-Abelian ${\rm su(2)}$ with
constant field strength,
\begin{eqnarray}\label{eq:Achoice}
  A_\mu^a(x)=\mathcal{A}_\mu(x) n^a\,,\quad {\rm with}\quad  
  \mathcal A_\mu=\frac{1}{2}x^\nu\mathcal{F}_{\nu\mu}\,,
\end{eqnarray} 
where $n^a$ is a constant unit vector in color space with $n^an^a=1$. The gauge fields \eq{eq:Achoice} satisfy the Fock-Schwinger gauge
$x^\mu A_\mu=0$. With \eq{eq:Achoice} we rewrite the Lorentz matrices
$M$ and $N$ as 
\begin{eqnarray}
  M(x)&=&2i(\mathcal{F}_{\mu\nu}-\delta_{\mu\nu}\frac{1}{4}x^\rho
  \mathcal F_{\rho\gamma}\dot x^\gamma)\otimes\mathcal T_{-}\,,\nonumber\\
  N(x)&=&-\frac{i}{2}x^{\mu}\mathcal{F}_{\mu\nu}\dot x^\nu\otimes
  \mathcal T_{-}\,,
\end{eqnarray}
with $\mathcal T_{-}=n^a\lambda^a$. The computation of $\Gamma_{\rm
  ghost}$ is straightforward but tedious \cite{Zhang:2005}. It results in 
\begin{eqnarray}\nonumber 
  \hspace{-1.5cm}\Gamma_{\rm ghost}&=& \frac{1}{2 (4\pi)^{D}}
  \int_{\tau}^{\infty}
  dT_1dT_2dT_3\\
  && \hspace{1.3cm} \times 
  e^{-m^2 T} {\cal I}_{\rm ghost}[T_1,T_2,T_3;{\cal F}]+{\rm c.t.}\,,
 \label{eq:Gaghost}\end{eqnarray}
with $T=T_1+T_2+T_3$. The integrand ${\cal I}_{\rm ghost}$ of the
proper-time integral is given by 
\begin{eqnarray}\label{eq:Ighost}
  \hspace{-.6cm}{\cal I}_{\rm ghost}&=& 
  \int_y \Bigl(\frac{2}{T_3}\tr\left(\CF\cot \CF
    T_1 \,G^{-1}\,\cos 2\CF T_2\right)\\\nonumber
  &+&\tr\left(G^{-1} \CF^2\cot\CF T_1 
    \cot\CF T_2\right)\\\nonumber 
  &-&\frac{2}{T_3}\tr\left(\CF
    G^{-1}\sin 2\CF T_2\right)+\tr\left(\CF^2 G^{-1}
  \right)\Bigr) \det{}^{\s012} R\,,
\end{eqnarray}
where  
\begin{eqnarray}\label{eq:GR}
\hspace{-.4cm}  G&=& \CF\cot\CF T_2+\CF\cot\CF T_1+\frac{1}{T_3}\,,\\
\nonumber 
\hspace{-.4cm}  R&=& \frac{\CF^2}{\sin\CF T_1 
      \sin\CF T_2(1+\CF T_3(\cot \CF T_1+\cot \CF T_2))}\,.
\end{eqnarray}
The expression \eq{eq:Gaghost} with \eq{eq:Ighost} is numerically accessible, 
after the counter terms in \eq{eq:Gaghost} are specified. \smallstep

\section
{renormalisation}\label{sec:renorm}

Now we discuss the UV subtraction terms hidden in the counter terms that render $Z_2$, $\Gamma_2$ finite, and in particular,
\eq{eq:Gaghost} finite. Apart from applying a standard dimensional
regularisation convenient for analytic considerations, we have
introduced a gauge invariant UV regularisation by cutting off the
proper-time integrals in \eq{eq:Gaghost} at a finite proper time
$\tau$, $T_i\geq \tau$. This translates via a Laplace transform into a
gauge invariant momentum cut-off if the effective action is formulated
in terms of momentum loops. Such a scheme makes numerical computations
accessible where the dimensional regularisation only can be employed
in exceptional cases. Indeed, a fully non-perturbative wordline
formulation of quantum field theories would provide a tool for
devicing gauge-invariant momentum cut-off schemes on the
non-perturbative level which would be highly interesting.  The ghost
action \eq{eq:preGaghost} is then written as $\Gamma_{\rm
  ghost}=\Gamma_{\rm ghost,reg}+{\rm c.t.}$ with
\begin{eqnarray}\nonumber 
\Gamma_{\rm ghost,reg}&=& \frac{1}{2 (4\pi)^{D}}
\int_{\tau}^{\infty}
  dT_1dT_2dT_3\, e^{-m^2 T}\\
&& \hspace{1cm} \times 
 {\cal I}_{\rm ghost}[T_1,T_2,T_3;{\cal F}]\,. 
 \label{eq:Gghostreg} \end{eqnarray}

The regularised expression $\Gamma_{\rm ghost,reg}$ diverges if the
regularisation parameters $\tau,\epsilon$ are removed. Here we first
concentrate on the proper-time regularisation with $\epsilon=0$. Then
the regularised expression $\Gamma_{\rm ghost,reg}$ in
\eq{eq:Gghostreg} diverges with powers of $1/\tau$, more precisely
with $1/\tau^n (\ln\tau)^m$.  Moreover, since we are dealing with the
two-loop effective action, the divergent terms are not necessarily
polynomial, and the counter terms cannot be determined in a polynomial
expansion. The non-polynomial terms can be attributed to the
divergence of one loop sub-diagrams which can be used to construct the
related counter terms. However, here we want to set-up a procedure
with which these counter terms can be derived systematically from
$\Gamma_{\rm ghost,reg}$ by means of derivatives. Such a procedure
mimics the standard BPHZ-type subtraction schemes in momentum space.
The divergences in $\tau$ are extracted by appropriate
$\tau$-derivatives with the help of the identity
\begin{eqnarray}\label{eq:del-ids} 
  \tau\0{\partial}{\partial \tau} \tau^{-n} (\ln\tau)^m= 
  - \tau^{-n} 
  (\ln\tau)^m \left(n- \0{m}{\ln \tau}\right)\,.
\end{eqnarray}
Applying the above $\tau$-derivative to $\Gamma_{\rm ghost,reg}$, 
the physically finite term drops out. Hence appropriate subtractions 
\begin{eqnarray}\label{eq:finite} 
\Gamma=\Gamma_{\rm reg} -
\sum_n c_n(\tau)(\tau\partial_\tau )^n\Gamma_{\rm reg}\,,
\end{eqnarray}  
render the action $\Gamma$ finite. Note that such a procedure in
principle only properly provides the renormalised effective action by
a careful discussion of the finite renormalisation that originates in
the subtractions in \eq{eq:finite}. In the present two loop case it
suffices to only take one $\tau$-derivative, e.g.\
\begin{eqnarray}\nonumber 
\tau \partial_\tau \Gamma_{\rm ghost,reg} &=& -\frac{\tau}{2(4\pi)^D}
\int_{\tau}^{\infty} dT_1 dT_2 d T_3 \sum_{i=1}^3 
\delta(T_i-\tau)   \\
&& \hspace{.3cm} \times 
 e^{-m^2 T} {\cal I}_{\rm ghost}[T_1,T_2,T_3;{\cal F}]\,. 
\label{eq:ep-der}
\end{eqnarray} 
This reduces the number of proper-time integrations and makes the
divergence structure analytically accessible. This procedure deserves
further studies.

We still have to compute the traces over the field strength.
Following \cite{Kors:1998ew}, we work in the Lorentz frame in which the
electric and magnetic fields are parallel, and thereby the field
strength takes on a simple form with only two non-zero symplectic
block elements, i.e., it can be written as
\begin{eqnarray}
\CF=a\sigma_1+b\sigma_2
\end{eqnarray}
with
\begin{eqnarray}
  \sigma_1 = \left(\begin{array}{cc} 
      \sigma & 0 \\ 0 & 0 \end{array} 
  \right),\;\;\;
  \sigma_2 = \left(\begin{array}{cc} 0 & 0 \\ 0 & 
      \sigma \end{array} 
  \right),\;\;
\end{eqnarray}
and
\begin{eqnarray}
a=\epsilon\,,\;\; b=-i \eta\,,
\end{eqnarray}
where $\epsilon$ and $\eta$ are the magnitudes of the magnetic and electric
fields, respectively.

We close with a remark on the explicit computation of the proper-time
integrals. In particular for numerical purposes it is advantageous to
convert the integrals into less divergent expressions. Indeed, ${ \cal
  I}_{\rm ghost}$ as well as the corresponding integrand ${ \cal
  I}_{\rm gluon}$ can be integrated analytically over $T_3$ and hence
can be written as a total $T_3$-derivative. The computations are
deferred to Appendix~\ref{app:ghost} and Appendix~\ref{app:gluon}
respectively and the results read
 \begin{eqnarray}\label{eq:T_3ghost} 
\Gamma_{\rm ghost}&=&\frac{1}{2 (4\pi)^{D}}\int_\tau^\infty 
  \prod_{i=1}^3 d T_i\, e^{-m^2 T} \\\nonumber 
  && \hspace{1.5cm}\times  
\partial_{T_3} \hat {\cal I}_{\rm ghost}[T_1,T_2,T_3;\CF]+{\rm c.t.}\,, 
\end{eqnarray} 
where $\hat {\cal I}_{\rm ghost}$ is given in \eq{eq:hatIghost},
 and  
 \begin{eqnarray}\label{eq:T_3gluon} 
\Gamma_{\rm gluon} &=&-\frac{1}{2 (4\pi)^{D}}\int_\tau^\infty 
  \prod_{i=1}^3 d T_i\, e^{-m^2 T}\\\nonumber 
  && \hspace{1.5cm}\times  
\partial_{T_3} \hat {\cal I}_{\rm gluon}[T_1,T_2,T_3;\CF]+{\rm c.t.}\,, 
\end{eqnarray} 
where $\hat {\cal I}_{\rm gluon}$ is given in \eq{eq:hatIgluon}. The
counter terms in \eq{eq:T_3ghost},\,\eq{eq:T_3gluon} are
$\tau$-dependent and can be computed from the derivative procedure
outlined in \eq{eq:del-ids},\,\eq{eq:finite}.

\section
{Two loop $\beta$ function}

In the remainder of this work we concentrate on the the question of
full two-loop consistency of the wordline formalism suggested in
\cite{Sato:1998sf,Sato:2000cr}, the construction of which we have
completed here.  To that end we discuss the running of the coupling at
two loop which is universal in mass-independent renormalisation
schemes. As this concerns an analytic computation we employ a
dimensional regularisation with $\tau=0$. Moreover, we use a minimal
subtraction scheme that renders the renormalisation constants
mass-independent, and hence projects onto the the universal result for
the two loop $\beta$-function. How such a mass-independent scheme is
fixed in the presence of general IR cutoffs has been discussed in
detail in \cite{Pawlowski:2005xe}, and we can straightforwardly use
the related arguments for the mass terms for gluon and ghosts employed
in the present work. The $\beta$-function can be read-off from the
running of the wave function renormalisation $Z_A$ of the gauge field.
Expanding the two loop contribution to the effective action
$\Gamma_{2,\rm reg}$ in powers of $\CF$ we are led to
\begin{eqnarray}\label{eq:Gaexp}
\Gamma_2[A]=\0{Z_A}{4} \int_x \tr\, \CF^2+O(\CF^3)\,. 
\end{eqnarray} 
The $\CF^2$-coefficient of the ghost effective action $\Gamma_{\rm
  ghost,reg}$ reads 
\begin{eqnarray}\label{eq:pole}
  &&\frac{4 g^4}{(4\pi)^4}
  \{-\frac{10}{3}C_1'-2C_2'-\frac{8}{3}C_3'+\frac{3}{2}C_4'\nonumber\\
&&-(\frac{1}{3}C_1'+\frac{1}{3}C_2'+\frac{5}{6}C_3'+C_4')\epsilon\} 
\0{1}{4}\int_x \tr\, \CF^2\,,
\end{eqnarray} 
where the coefficients $C_i'$ are given by \cite{Sato:2000cr}
\begin{eqnarray}
  C_1'\!\!&=&\!\!(4\pi\mu^2)^{2\epsilon}
  \int_0^{\infty}\prod_{i=1}^3 dT_i\,\0{T_1^4T_2}{\Omega^{4-\epsilon}}\,
  e^{-m^2T}\,,\nonumber\\
  C_2'\!\!&=&\!\!(4\pi\mu^2)^{2\epsilon}
  \int_0^{\infty}\prod_{i=1}^3 dT_i\,\0{T_1^3
    T_2^2}{\Omega^{4-\epsilon}}\,e^{-m^2T}\,,\nonumber\\
  C_3'\!\!&=&\!\!(4\pi\mu^2)^{2\epsilon}
  \int_0^{\infty}\prod_{i=1}^3 dT_i\,\0{T_1^3
    T_2T_3}{\Omega^{4-\epsilon}}\,e^{-m^2T}\,,\nonumber\\
  C_4'\!\!&=&\!\!(4\pi\mu^2)^{2\epsilon}
  \int_0^{\infty}\prod_{i=1}^3 dT_i\,\0{T_1^2T_2^2T_3}{\Omega^{4-\epsilon}}
  \,e^{-m^2T}\,,
\label{eq:coeffs}\end{eqnarray}
with $\Omega=T_1T_2+T_2T_3+T_3T_1$. The divergent
parts of the coefficients $C_i'$ read
\begin{eqnarray}
  C_1'&=&-\frac{1}{6\epsilon^2}+(-\frac{5}{9}+\frac{\rho_m}{3})
  \frac{1}{\epsilon}+\mathcal
  O(\epsilon^0)\,,\nonumber\\
  C_2'&=&\frac{1}{6\epsilon^2}+(\frac{1}{18}-\frac{\rho_m}{3})
  \frac{1}{\epsilon}+\mathcal
  O(\epsilon^0)\,,\nonumber\\
  C_3'&=&\frac{1}{12\epsilon^2}-(\frac{1}{72}+\frac{\rho_m}{6})
  \frac{1}{\epsilon}+\mathcal
  O(\epsilon^0)\,,\nonumber\\
  C_4'&=&\frac{1}{12\epsilon}+\mathcal O(\epsilon^0)\label{eq:C's}\,,
\end{eqnarray}
with
\begin{equation}
  \rho_m=\gamma_E+\mbox{ln}\frac{m^2}{4\pi\mu^2},\;\;\;\;\;\;
  \gamma_E=\mbox{Euler
    const}\,.
\end{equation}
Now we are in the position to perform the UV renormalisation for the
ghost term. Note that by power counting, the UV divergence can appear
at most in the quadratic term in the expansion, we write
\begin{eqnarray}\label{eq:renormexpl}
  \hspace{-.8cm}\Gamma_{\rm ghost,ren}
  &=&\Gamma_{\rm ghost}-\014 C'_{\rm ghost}\int_x \tr \CF^2 \nonumber\\
&+&  \014 C'_{\rm finite}\int_x \tr \CF^2\,,
\end{eqnarray} 
where
\begin{eqnarray}\label{eq:coeffC}
&&C'_{\rm ghost}=\frac{4 g^4}{(4\pi)^4}\{-\frac{10}{3}C_1'-2C_2'-\frac{8}{3}C_3'
+\frac{3}{2}C_4'\nonumber\\
&&-(\frac{1}{3}C_1'+\frac{1}{3}C_2'+\frac{5}{6}C_3'+C_4')\epsilon\}\,,
\end{eqnarray}
and $ C'_{\rm finite}$ is its finite part. This introduces the minimal
subtraction scheme in the ghost part. The integral in \eq{eq:Gaghost}
is changed by the additional integrands proportional to $C'_{\rm ghost}\CF^2$
rendering a finite expression. We remark that it can be explicitly
checked that the renormalisation constants are mass-independent at two
loop. This constitutes a mass-independent RG-scheme and hence
$\beta_2$ is universal. \smallstep

The two loop
$\beta$-function is provided by 
\begin{eqnarray}\label{eq:beta}
\beta&=&-g\,\mu\,\partial_\mu \ln Z_g\\ \nonumber 
&=&-g \left(\beta_1 C_A \left(\0{g}{4 \pi} 
\right)^2+\beta_2 C_A^2 \left(\0{g}{4 \pi}\right)^4+O(g^6)\right)\,.
\end{eqnarray}
 The background field formalism allows us to directly
extract the two loop $\beta$-function from $Z_A$: the effective action
$\Gamma[A]$ is gauge invariant and consequently the combination $gA$
is RG-invariant, leading to $Z_g=Z_A^{-1/2}$, and hence 
\begin{eqnarray}\label{eq:betaZA}
\beta=\012 g\,\mu\,\partial_\mu \ln Z_A\,.
\end{eqnarray}
With \eq{eq:beta} and \eq{eq:betaZA} we conclude that
\begin{eqnarray}
  Z_A=1+\frac{\beta_1}{\epsilon}C_A \frac{g^2}{(4\pi)^2}
  +\frac{\beta_2}{\epsilon}C_A^2\frac{g^4}{(4\pi)^4}+O(g^6)\,,  
\end{eqnarray}
and we directly read off 
the two loop $\beta$-function from the subtraction terms computed in the last 
section. Using \eq{eq:C's} in \eq{eq:pole} we arrive at the
ghost contribution $\beta_{2,\rm ghost}$ to the two loop coefficient
$\beta_2$,
\begin{eqnarray}\label{eq:bghost}
  \beta_{2,\rm ghost}=\0{11}{6}\,.
\end{eqnarray}
The gluon loop contribution has been computed in \cite{Sato:2000cr} as
$\beta_{2,\rm gluon}=-11/2$. It is left to compute the contributions of 
the one loop counter terms. They arise from the insertion of the one loop RG 
constants of coupling and propagating field at one loop.  Note
that the propagating field is the fluctuation field $a$ with one loop
wave function renormalisation 
\begin{eqnarray}\label{eq:wave}
  \delta Z_a= \0{1}{\epsilon}\frac{g^2}{(4\pi)^2}
  \left(\0{5}{3}+\0{1}{2}(1-\xi)\right)\,.
\end{eqnarray}
The worldline counter terms reduce to the standard one loop graphs.
The corresponding diagrams shown in Fig.1 and 2 result from the
one-loop renormalisation of the fluctuation field $a$ and its
vertices, while those in Fig.3 and 4 arise from the one-loop
renormalisation of ghost field and its vertices.
\begin{figure}[htbp]
\centering
\includegraphics[width=.4\textwidth]{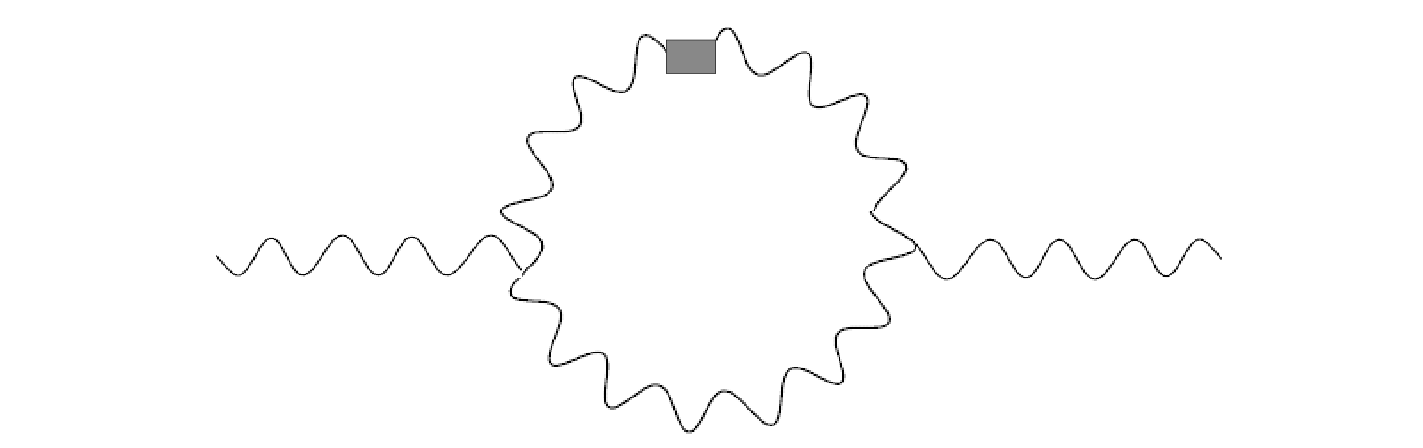}
\ \vspace{.3cm} \
\caption{counter terms from the propagator renormalisation of the
  fluctuation field $a$\,.}
\end{figure}
\begin{figure}[htbp]
\centering
\includegraphics[width=.4\textwidth]{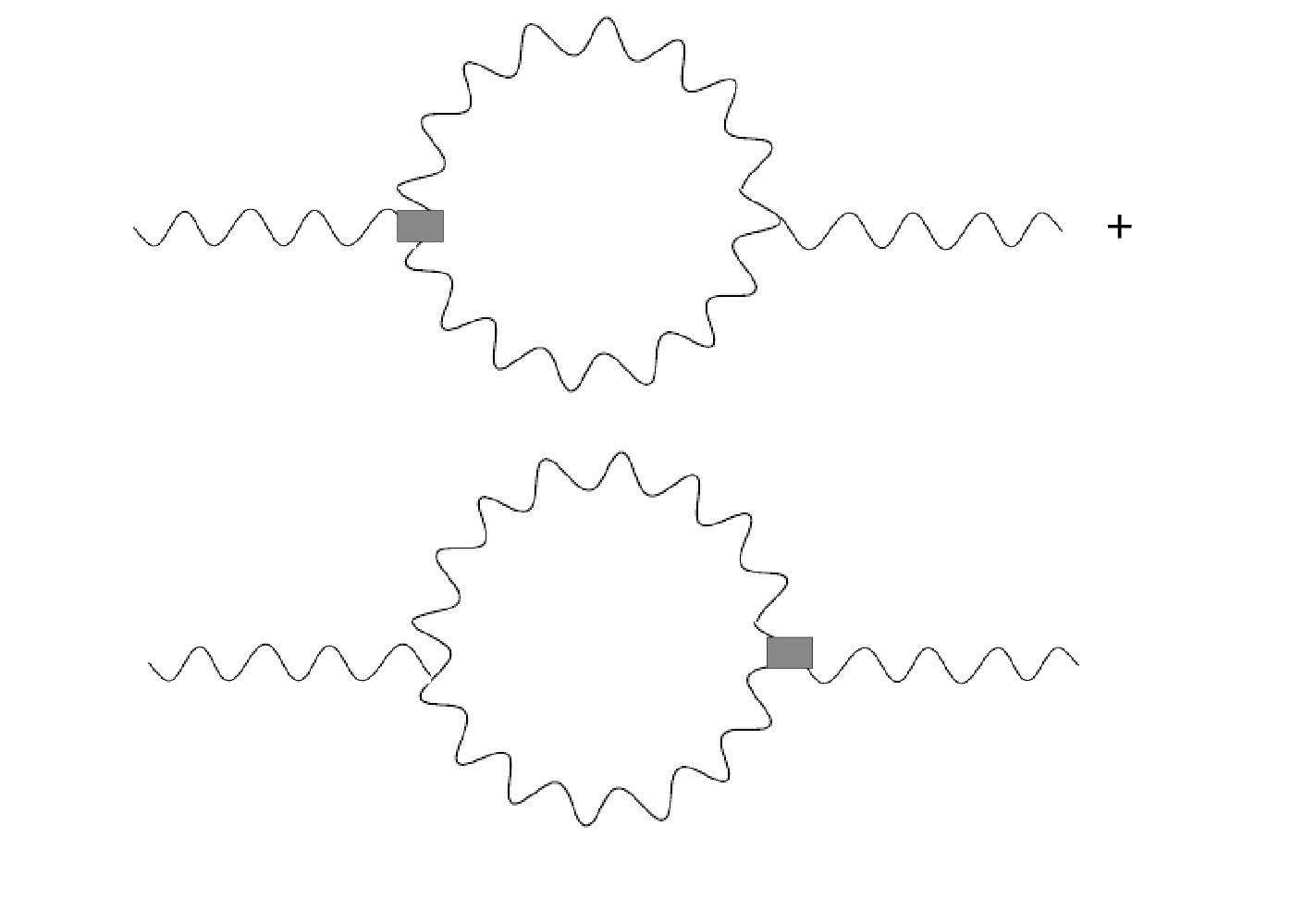}
\caption{counter terms from the vertex renormalisation\,.}
\end{figure}
\begin{figure}[htbp]
  \centering
\includegraphics[width=.4\textwidth]{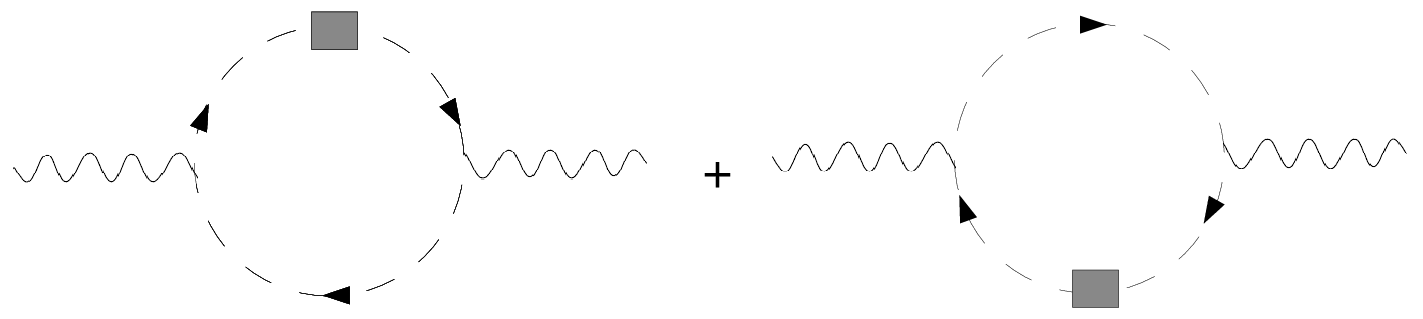}
\caption{counter terms from the ghost propagator renormalisation\,.}
\centering
\vspace{1em}
\includegraphics[width=.4\textwidth]{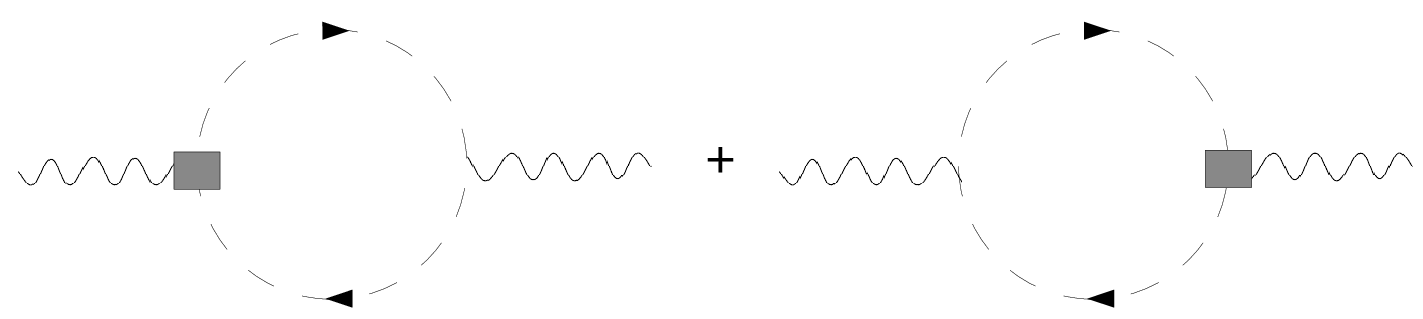}
\caption{counter terms from the vertex renormalisation\,.}
\end{figure}

The computation of the counter term in Fig.1 requires a gluon mass
renormalisation with $m^2\to Z_a Z_m m^2$. In Feynman gauge it is
given by \cite{Chetyrkin:1997fm}
\begin{eqnarray}\label{eq:massren}
 Z_a  Z_m=1-\0{1}{\epsilon}\frac{g^2 C_A}{(4\pi)^2}\,.
\end{eqnarray}
This counter term has been considered in \cite{Sato:2000cr},
which gives a contribution of $10/3\epsilon$ to the two loop
coefficient $\beta_2$. We also remark that the gauge-fixing term does
not renormalise, $\xi \to Z_a \xi$.

\noindent 
Note that the ghost mass term is not renormalised. The counter terms from
Fig.1-Fig.4 give rise to pole contributions
\begin{eqnarray}
  &\mbox{Fig.1 + Fig.2}: &\hspace{1cm} \frac{35}{6\epsilon}\,,\nonumber\\
  &\mbox{Fig.3 + Fig.4}:&\hspace{1cm} \frac{1}{6\epsilon}\,,
\end{eqnarray}
and we are led to
\begin{equation}
  \beta_2=-\frac{11}{2}
  +\frac{11}{6}+\frac{35}{6}+\frac{10}{3}+\frac{1}{6}=\frac{17}{3}\,,
\end{equation}
which agrees with the well-known result, e.g.\
\cite{Caswell:1974gg,Abbott:1980hw}.  As we are only interested in
diagrams with external background fields we could have rescaled the
fluctuation field $a$ and the ghost field with the renormalisation
factors. With these rescaled fields, the diagrams above reduce to
terms proportional to the renormalisation of the gauge-fixing term
introduced by this rescaling and the renormalisation of the mass terms
which also changes by this rescaling (e.g.\ the ghost mass
renormalises with these rescaled fields). This has been used in
\cite{Abbott:1980hw}. Of course, this does not change the result. For
comparison we list the different contributions \ \vspace{.3cm} \
$$\begin{tabular}{clcrc} \hline\bfseries{$\beta_2$-contributions}
  &\multicolumn{3}{c} {$\bf m^2\neq 0$} &{
    $\bf m^2=0$} \\\hline\\
  two-loop diagrams&\multicolumn{3}{c} {$-\frac{11}{3\epsilon}$}
  & $\frac{7}{3\epsilon}$\\\\
  Fig.1+2 &$\frac{5}{18\epsilon}+\frac{80}{9\epsilon}$ & or &
  $\frac{35}{6\epsilon}+\frac{10}{3\epsilon}$ &
  $\frac{10}{3\epsilon}$\\\\
  Fig.3+4 & $0+\frac{1}{6\epsilon}$& or
  &$\frac{1}{6\epsilon}+0$  &0\\\\
  total &\multicolumn{3}{c}{$\frac{17}{3\epsilon}$}
  &$\frac{17}{3\epsilon}$\\\\\hline
\end{tabular}$$
\ 
\vspace{.3cm}
\ 

In the middle column the different contributions from a direct
computation (right), and from one with rescaled fluctuation fields
(left) are listed.  \smallstep

We can use the above results on the consistent renormalisation in the
presence of an infrared mass-scale to define the renormalised two loop
contribution $\Gamma_2[A]$ by means of appropriate subtractions
instead of the dimensional regularisation used above. This allows us 
to numerically compute the full two loop effective action
$\Gamma=\Gamma_1+\Gamma_2$ as a function of $\CF$.

\section{outlook}
In the present work we have completed the wordline construction of the
two loop effective action initiated in \cite{Sato:1998sf,Sato:2000cr}.
In particular we have provided a crucial consistency check of the
construction by computing the universal two loop $\beta$-function
within the wordline formalism.

We also have set-up a practical ultraviolet BPHZ-type renormalisation
scheme in the proper-time which makes numerical computations accessible. 
For example, this can be used to numerically compute the two loop effective 
action for covariantly constant fields. 

The inclusion of fermions in the present approach is straightforward,
and is, in our opinion, the physically most interesting extension of
the present work.
\\

\noindent {\it Acknowledgements} We thank G.~Dunne, H.~Gies and 
C.~Schubert for useful discussions. 

\setcounter{section}{0}
\setcounter{equation}{0}
\renewcommand{\thesection}{\Alph{section}}
\renewcommand{\thesubsection}{\Alph{section}}
\renewcommand{\theequation}{\Alph{section}.\arabic{equation}}

\section
{Ghost contribution}\label{app:ghost} 

One of the $T$-integrations in \eq{eq:Gaghost} can be done
analytically.  This is achieved by integrating ${\cal I}_{\rm ghost}$
over $T_3$.  Performing all the traces in the integrand of \eq{eq:Ighost}
and summing over them gives 
\begin{eqnarray}\label{eq:A}
  A&=&\frac{2a}{1+a a_1 T_3}\Big(-a T_3+2\cosh(2a T_2)\coth(a T_1)
  \nonumber\\
  &+&a T_3\coth(a T_1)\coth(a T_2)+2\sinh(2a T_2)\Big)\nonumber\\
  &+&(a\leftrightarrow b)\,, 
\end{eqnarray} 
with $a_1=\coth(aT_1)+\coth(aT_2)$. The square-rooted determinant term
reads
\begin{eqnarray}\label{eq:det} 
  B=\frac{C_1}{(1+ a a_1 T_3)(1+b
    b_1 T_3)}\,,
\end{eqnarray} 
where $C_1=a^2b^2\,\mbox{csch}(aT_1)\,\mbox{csch}(bT_1)\,
\mbox{csch}(aT_2)\,\mbox{csch}(bT_2)$. We define 
\begin{equation}
  \hat{\cal I}_{\rm ghost}[T_1,T_2,T_3;{\cal F}]=
\int^{T_3}_0 dT'_3\,{\cal I}_{\rm ghost}[T_1,T_2,T'_3;{\cal F}]\,,
\end{equation} 
after analytically performing the integration over $T_3$
one finds ($b_1=a_1(a\leftrightarrow b)$)
\begin{eqnarray} \label{eq:hatIghost}
\hat{\cal I}_{\rm ghost}&=&\int_y\Big\{\Big[\frac{A C_1
    \left(\ln(1+a a_1 T_3)-\ln(1+b b_1 T_3)\right)}{a a_1-b b_1}
  \nonumber\\
  &-&\Big(\frac{C_1C_2}{a a_1(a a_1-b b_1)^2(1+a a_1 T_3)}
  \big(-a a_1+b b_1\nonumber\\
  &-&a a_1(1+b b_1 T_3)\ln(1+a a_1 T_3)\nonumber\\
  &+&a a_1(1+b b_1 T_3)\ln(1+b b_1 T_3)\big)\nonumber\\
  &+&(a\leftrightarrow b)\Big)\Big]-[T_3\to0]\Big\}\,,
\end{eqnarray} 
with
\begin{eqnarray}
  \hspace{-0.4cm}   C_2&=&2a^2\Big(-1-2a_1\cosh(2a T_2)\coth(a T_1)
  \nonumber\\
  &+&\coth(a T_1)\coth(a T_2)-2a_1\sinh(2a T_2)\Big)\,.
\end{eqnarray}

\section
{Gluonic contribution}\label{app:gluon} 
The gluon loop contribution to the two loop effective action reads
\cite{Sato:2000cr}
\begin{eqnarray}\nonumber 
  &&\hspace{-.3cm} \Gamma_{\rm gluon}=
-\frac{1}{2}(4\pi)^{-D}\int_0^{\infty} dT_1dT_2dT_3\, 
e^{-m^2 T}{\cal I}_{\rm gluon}+{\rm c.t.}\\\nonumber
  & &=-\frac{1}{2}(4\pi)^{-D}\int_0^{\infty} dT_1dT_2dT_3\, 
e^{-m^2 T} \int_y\,\det{}^{\frac{1}{2}}(\frac{\mathcal F^2}{
    \Delta_{\mathcal F}})\\\nonumber
  &&
\Bigl[\tr\Bigl(\frac{\mathcal F^2 T_3}{\Delta_\mathcal F
    \sin {\cal F}T_2} \Bigl(2\sin {\cal F}T_1\, 
  \cos 2{\cal F}(T_1+2T_2)  \\\nonumber 
\end{eqnarray}
\begin{eqnarray}\nonumber
  & &-2\sin {\cal F}(T_1+T_2)\, \cos{\cal F}(2T_1+3T_2)+\Bigl[1
\\\nonumber 
  &&
-2\cos 2{\cal F}(T_1+T_2)\Bigr] 
  \sin({\cal F}T_2)\cos{\cal F}(T_1-T_2)\Bigr)\\\nonumber 
  & &+\frac{{\cal F}}{\Delta_{\cal F}}[4\sin{\cal F}T_1\, 
  \sin {\cal F}T_2\,
  \sin 2{\cal F}(T_1+T_2)
\\\nonumber 
  &&
-2\sin {\cal F}T_1\,\cos 
  {\cal F}(2T_1+3T_2)
\\\nonumber 
  && 
-2\sin {\cal F}T_2\,\cos{\cal F}(T_1-2T_2)
\\\nonumber 
  &&
  -\sin 
  {\cal F}(T_1+T_2)\,\cos 2{\cal F}(T_1-T_2)]  \Bigr)\\\nonumber
  &&+\tr\,\cos 2{\cal F}T_2\,
  \tr\, \Bigl[\frac{{\cal F}^2 T_3}{\Delta_{\cal F}\sin {\cal F}T_2 }
  \Bigl(\sin {\cal F}(T_1+T_2)
  \\\nonumber
  & & 
\times\cos{\cal F}(2T_1+T_2)-
  \sin {\cal F}T_1\,\cos 2{\cal F}(T_1+T_2)\Bigr)
\\\nonumber
 & &
+\frac{{\cal F}}{\Delta_{\cal F}}\Bigl(3
  \sin {\cal F}T_1\,\cos {\cal F}(2T_1+T_2)\\\nonumber 
  &&  +\cos 2{\cal F}T_1\,\sin {\cal F}(T_1+T_2)\Bigr)\Bigl]\\\nonumber
  &&  +\tr \Bigl(\frac{{\cal F}^2 T_3}{\Delta_{\cal F}
    \sin {\cal F}T_2 }\Bigl[\sin{\cal F}T_2\,\cos 
  {\cal F}(T_1-T_2)
\\\nonumber 
  &&
+\cos {\cal F}T_2\,\sin{\cal F}(T_1+T_2)\\\nonumber 
  &&-\sin{\cal F}T_1\,\cos 2{\cal F}T_2\Bigr] +
  \frac{{\cal F}}{\Delta_{\cal F}}\sin {\cal F}T_1\,
  \cos {\cal F}T_2 \Bigr)
 \\\nonumber 
  &&
\times \tr\Bigl(
  \cos 2{\cal F}(T_1+T_2) \Bigr)
  +\delta(T_2)2(1-D)\tr\Bigl(\cos 2{\cal F}T_1\Bigr)\nonumber\\
  &&+
  \delta(T_3)\tr\Bigl(\cos 2{\cal F}(T_1-T_2)\Bigl)\nonumber\\
  &&
-\delta(T_3)\tr\Bigl(\cos 2{\cal F}T_1\Bigr) 
  \tr\Bigl(\cos 2{\cal F}T_2\Bigr)\Bigr]\nonumber\\
  &&+\frac{1}{2}(4\pi)^{-D}\int_0^{\infty}dT_1dT_2\int_y\,
  \mbox{det}^{\frac{1}{2}} \frac{{\cal F}^2}{\sin {\cal F}T_1 
    \,\sin {\cal F}T_2 } 
\nonumber\\
  &&
\Bigl(\tr\cos 2{\cal F}(T_1+T_2) -
  \tr \cos 2{\cal F}(T_1-T_2)\Bigr)+{\rm c.t.}
\label{eq:Gagluon}\end{eqnarray}  
with
\begin{equation}
  \Delta_{\cal F}=\sin({\cal F}T_1)\sin({\cal F}T_2)+{\cal F}T_3
  \sin[{\cal F}(T_1+T_2)]\,.
\end{equation} 
Following the same procedure as we extracted Eq.~(\ref{eq:pole}) from
(\ref{eq:Gaghost}), we obtain the renormalisation part of the
effective action above at the second order of $\mathcal F$
\begin{equation}
  \Gamma_{\rm gluon}=\frac{4g_0^4}{(4\pi)^4}(-\frac{11}{2\epsilon})
  \int_y(-\frac{1}{4}{\cal F}_{\mu\nu}{
    \cal F}_{\mu\nu})+\mathcal O(\epsilon^0)\,,
\end{equation} 
and hence the contribution of the gluon loops to the two loop
$\beta$-function coefficient is $-11/2\epsilon$. \smallstep

For a numerical evaluation of $\Gamma_{\rm gluon}$ a less singular
representation is advantageous. To that end we write $\Gamma_{\rm
  gluon}$ as
\begin{eqnarray}\label{eq:T_3-dergluon} 
\Gamma_{\rm gluon} &=&-\frac{1}{2 (4\pi)^{D}}\int_\tau^\infty 
  \prod_{i=1}^3 d T_i\, e^{-m^2 T}\\\nonumber 
  && \hspace{1.5cm}\times  
\partial_{T_3} \hat {\cal I}_{\rm gluon}[T_1,T_2,T_3;\CF]+{\rm c.t.}\,, 
\end{eqnarray} 
with $\partial_{T_3} \hat {\cal I}_{\rm gluon}= {\cal I}_{\rm gluon}$, and 
hence similarly to the ghost-contribution it reads 
\begin{equation}
  \hat{\cal I}_{\rm gluon}[T_1,T_2,T_3;{\cal F}]=
\int^{T_3}_0 dT'_3\,{\cal I}_{\rm gluon}[T_1,T_2,T'_3;{\cal F}]\,,
\end{equation} 
The $T_3$-integral in the definition of $\hat {\cal
  I}_{\rm gluon}$ can be performed analytically and yields 
\begin{widetext}
\begin{eqnarray}
  \hat {\cal I}_{\rm gluon}&=&\int_y\Big\{\Big[\frac{A' C_1
    \left(\ln(1+a a_1 T_3)-\ln(1+b b_1 T_3)\right)}{a a_1-b b_1}+\Big(C_1 C_3\,\theta(T_3)+\frac{2a C_1 
    (a_1 C_5-C_4)}{a_1(a a_1-b b_1)^2(1+a a_1 T_3)}\nonumber\\
  &\times&\big(-a a_1+b b_1-a a_1(1+b b_1 T_3)\ln(1+a a_1 T_3)+a a_1(1+b b_1 T_3)\ln(1+b b_1 T_3)\big)+(a\leftrightarrow b)\Big)\Big]-[T_3\to0]\Big\}\,,\nonumber\\
\label{eq:hatIgluon}\end{eqnarray}
\end{widetext}
with $\theta(T_3)$ the step function and
\begin{widetext}
\begin{eqnarray}
  A'&=&\frac{2a (C_5+C_4\,a T_3)}{1+a a_1 T_3}-4(-1+D)\cosh(2a T_1)\,
  \delta(T_2)
  +2\,\delta(T_3)\Big(\cosh(2a(T_1-T_2))-2\big(\cosh(2a T_1)\nonumber\\
  &+&\cosh(2b T_1)\big)\cosh(2a T_2)-2\sinh(2a T_1)\sinh(2a T_2)\Big)
  +(a\leftrightarrow b)\,.
\end{eqnarray}
The abbreviations $C_3$, $C_4$ and $C_5$ read 
\begin{eqnarray}
  C_3&=&2\cosh\big(2a(T_1-T_2)\big)-4\sinh(2a T_1)\sinh(2a T_2) 
  - 4\big(\cosh(2a T_1)+\cosh(2b T_1)\big)\cosh(2a T_2)\,,\nonumber
\end{eqnarray}
\begin{eqnarray}\nonumber
  C_4&=&4\cosh\big(2b(T_1+T_2)\big)\big(-1+\coth(aT_1)\coth(aT_2)\big)
+\Bigl(2\cosh(2bT_2)\cosh\big(a(T_1+T_2)\big)\nonumber\\
&+&2\cosh\big(a(T_1-T_2)\big)+\cosh\big(a(3T_1+T_2)\big)\Bigr)
\mbox{csch}(a T_1)\mbox{csch}(a T_2)\,,\nonumber
\end{eqnarray}
and 
\begin{eqnarray}
  C_5&=&2\cosh\big(2a(T_1+T_2)\big)\coth(a T_2)+2\cosh\big(2b(T_1+T_2)\big)
  \coth(a T_2)-2\cosh\big(a(T_1-2T_2)\big)\mbox{csch}(a T_1)\nonumber\\
  &+&6\cosh\big(a(2T_1+T_2)\big)\mbox{csch}(a T_2)\big(\cosh(2a T_2)+
  \cosh(2b T_2)\big)-2\cosh\big(a(2T_1+3T_2)\big)\mbox{csch}(a T_2)\nonumber\\
  &+&\mbox{csch}(a T_1)\mbox{csch}(a T_2)\sinh\big(a(T_1+T_2)
  \Big(2\cosh(2a T_1)\cosh(2a T_2)+2\cosh(2a T_1)\cosh(2b T_2)\nonumber\\
  &-&\cosh\big(2a(T_1-T_2)\big)\Big)-4\sinh\big(2a(T_1+T_2)\big)\,.
\end{eqnarray}
\end{widetext}

\end{document}